\documentclass[10pt,aps,prb, twocolumn]{revtex4} 

\usepackage{amsmath} 
\usepackage{graphicx} 

\begin{document}

\title{Why is surface tension a force parallel to the interface?}

\author{Antonin Marchand$^1$}
\author{Joost H. Weijs$^2$}
\author{Jacco H. Snoeijer$^2$}
\author{Bruno Andreotti$^1$}

\affiliation{$^1$Physique et M\'ecanique des Milieux H\'et\'erog\`enes, UMR
7636 ESPCI -CNRS, Univ. Paris-Diderot, 10 rue Vauquelin, 75005, Paris,
France}
\affiliation{$^2$Physics of Fluids Group and J. M. Burgers Centre for Fluid Dynamics,
University of Twente, P.O. Box 217, 7500 AE Enschede, The Netherlands}
\date{\today}

\begin{abstract}
A paperclip can float on water. Drops of mercury refuse to spread on a surface. These capillary phenomena are macroscopic manifestations of molecular interactions, and can be explained in terms of surface tension. For students, the concept of surface tension is quite challenging since the microscopic intuition is often in conflict with the common macroscopic interpretations. In this paper we address a number of conceptual questions that are often encountered when teaching capillarity. By answering these questions we provide a perspective that reconciles the macroscopic viewpoints, from thermodynamics or fluid mechanics, and the microscopic perspective from statistical physics. 
\end{abstract}

\maketitle

\section{Basic concepts and problems}

Capillarity is one of the most interesting subjects to teach in condensed matter physics, as its detailed understanding involves three otherwise separated domains: macroscopic thermodynamics\cite{deGennes,Quere,Bonn09}, fluid mechanics and statistical physics\cite{KB1949}. The microscopic origin of surface tension lies in the intermolecular interactions and thermal effects\cite{Rowlinson/Widom,Berry}, while macroscopically it can be seen as a force acting along the interface or an energy per unit surface. In the present article we discuss the link between these three aspects of capillarity, on the basis of simple academic examples. We first discuss the standard problems faced by students --~and many researchers~-- in the understanding of surface tension. We will see that the difficulty of understanding surface tension forces is often caused by the improper or incomplete definition of a system on which the forces act. We bring up four basic questions, such as the one raised in the title, which are answered in the rest of the article. Contrary to many textbooks on the subject, this provides a picture that reconciles the microscopic, thermodynamic and mechanical aspects of capillarity. 

\subsection{The interface}
\subsubsection{Thermodynamic point of view}
\begin{figure}[t]
\begin{center}
\includegraphics{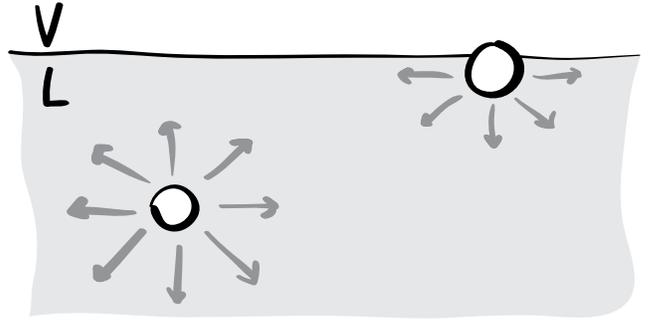}
\caption{\label{MissingBounds}Sketch showing the `missing' intermolecular bonds close to the liquid-vapor interface. This gives rise to an increase in free energy per unit area, i.e. the surface tension.}
\end{center}
\end{figure}

Following the pioneering work of Gibbs\cite{Gibbs}, we introduce surface tension as the excess free energy due to the presence of an interface between bulk phases. Let us consider a molecule in the vicinity of an interface, for example near a liquid-vapor interface. The environment of this molecule is manifestly different from the molecules in the bulk. This is usually represented schematically by drawing the attractive bonds around molecules, as shown in Fig.~\ref{MissingBounds}. One clearly sees from this picture that approximately half of the bonds are missing for a molecule at the interface, leading to an increase of the free energy. One thus defines the surface tension from the free energy $F$ per unit area: 
\begin{equation}
\gamma_{\rm LV}=\left( \frac{\partial F}{\partial A}\right)_{T,V,n},
\end{equation}
for a system of volume $V$ containing $n$ molecules at temperature $T$. Hence, $\gamma_{\rm LV}$ is the energy needed to increase the interfacial area by one unit. Its dimension is $[\gamma_{\rm LV}]=M.T^{-2}$ (mass per time squared) and it is usually expressed in $N/m$ (force per unit length) or $J/m^2$ (energy per unit area). 

It is instructive to estimate the magnitude of surface tension, which must be of the order of the bond energy $\epsilon$ divided by the cross section area $\sigma^2$ of a molecule --~$\sigma$ is a fraction of nanometer. For oils, interaction through van der Waals interactions leads to $\epsilon \sim k_BT\simeq\frac{1}{40}\,eV$ and thus $\gamma_{\rm LV}\sim 0.02$~N/m. For water, hydrogen bonds lead to a higher value $\gamma_{\rm LV}\sim0.072$~N/m. For mercury, the high energy bonds ($\epsilon \sim 1\,eV$) lead to a high surface tension $\gamma_{\rm LV}\sim 0.5$~N/m.

\subsubsection{Mechanical point of view}
\begin{figure}[t]
\begin{center}
\includegraphics{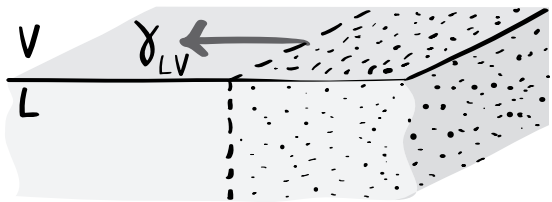}
\caption{\label{ForcePicture}Sketch showing surface tension as a force per unit length exerted by one subsystem on the other. The system on which the forces act is the dotted region. The force is parallel to the interface and perpendicular to the dividing line.}
\end{center}
\end{figure}

In fluid mechanics, the surface tension is not defined in terms of a surface energy but rather as a force per unit length. In the bulk of a fluid at rest, two sub-parts of a fluid exert a repulsive interaction on one another, which is called the pressure. If the surface separating these two subsystems crosses the liquid-vapor interface, an additional force needs to be taken into account: surface tension. As shown in Fig.~\ref{ForcePicture}, the surface tension is a force tangent to the surface and normal to the contour separating the two subsystems. The total force is proportional to the width of the contour, which we will call $W$ throughout the paper. Contrarily to pressure, surface tension is an attractive force.

The link between mechanics and thermodynamics is provided by the virtual work principle. If one moves a contour of width $W$ by a length $dl$, the area of the interface of the subsystem considered increases by $Wdl$. Consequently, the free energy is increased by $\gamma_{\rm LV} W dl$. The free energy should equal the work done by the surface tension force, which means that this force must be parallel to the interface, normal to the contour and have a magnitude $\gamma_{\rm LV} W$. Per unit length, the tension force is thus $\gamma_{\rm LV}$.

On the other hand, the link between mechanics and statistical physics is much less obvious for students. One clearly sees in Fig.~\ref{MissingBounds} that the molecule at the interface is submitted to a net force (which would be represented by the sum of the vectors) along the direction perpendicular to the interface. However, we just argued from the mechanical point of view, that the force is \emph{parallel} to the interface. This leads to the first key question of this article:\\
{\it $\bullet$ Q1: Why is surface tension a force parallel to the interface while it is so obvious that it must be perpendicular to it?}

\subsection{The contact line}\label{sec.plate}
\subsubsection{Thermodynamic point of view}
\begin{figure}[t]
\begin{center}
\includegraphics{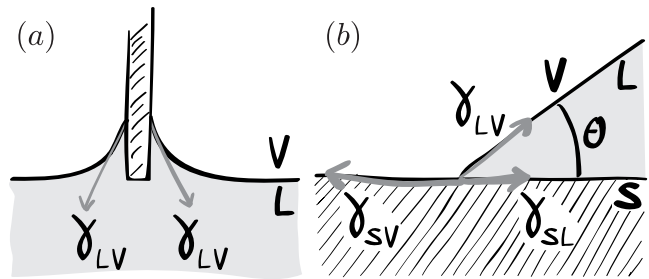}
\caption{\label{Young} (a) Experimental method to determine the liquid-vapor surface tension. The force per unit length needed to pull a plate from a bath of liquid is equal to $\gamma_{\rm LV} \cos \theta$, where $\theta$ is the equilibrium contact angle. (b) A classical way to interpret Young's law as a force balance of surface tensions. Question 2: Why is there no force balance in the normal direction? Question 3: Why do we draw a single surface tension force in the left panel ($\gamma_{\rm LV}$) while there are three in the right panel ($\gamma_{\rm LV},\gamma_{\rm SV}, \gamma_{\rm SL}$)?} 
\end{center}
\end{figure}

A standard experimental method for determining the liquid-vapor surface tension is to measure the force required to pull a metallic plate (usually made of platinum) out of a liquid bath. This force is related to the liquid-vapor surface tension $\gamma_{\rm LV}$, as is usually explained by the diagram of Fig.~\ref{Young}a. Imagine that the plate is moved vertically by a distance $dl$. The area of the liquid-vapor interface does not change by this motion, so the corresponding interfacial energy is unaffected. However, the motion does lead to a decrease of the immersed solid-liquid interface area, by $W\,dl$, while the solid-vapor interface increases by the same amount. In other words, part of the wetted surface is exchanged for dry surface. This leads to a variation of the free energy $dF=(\gamma_{\rm SV}-\gamma_{\rm SL})W\,dl$, where $\gamma_{\rm SV}$ and $\gamma_{\rm SL}$ are the solid-vapor and solid-liquid surface tensions respectively. This energy is provided by the work done by the experimentalist, due to the force required to displace the plate by $dl$. Hence, this force must be equal to $(\gamma_{\rm SV}-\gamma_{\rm SL})W$. 

To relate this force to the value of the liquid-vapor surface tension $\gamma_{\rm LV}$, one has to invoke Young's law for the contact angle $\theta$. When the three interfaces (between solid, liquid and vapor) join at the contact line, the liquid makes contact with the substrate at an angle $\theta$ given by\cite{Young}:
\begin{equation}
\label{young}
\gamma_{\rm LV} \cos \theta=\gamma_{\rm SV}-\gamma_{\rm SL}.
\end{equation}
With this, the force exerted on the plate can be expressed as $W\,\gamma_{\rm LV} \cos \theta$, and can thus be used to design a tensiometer.

\subsubsection{Mechanical point of view} 
From the mechanical point of view, we can interpret this force as the surface tension acting parallel to the liquid-vapor interface. By symmetry, the total force exerted on the solid is vertical (the horizontal components add up to zero). Projecting the surface tension force on the vertical direction and multiplying by the length $W$ of the contact line, one indeed gets $W~\gamma_{\rm LV} \cos \theta$. 

By a similar argument one usually interprets Young's law for the contact angle as the balance of forces at the contact line (Fig.~\ref{Young}b). By projection along the direction parallel to the solid substrate, one obtains $\gamma_{\rm SL}+\gamma_{\rm LV} \cos \theta=\gamma_{\rm SV}$, which is the same as equation~(\ref{young}). This force interpretation is a common source of confusion for students: \\
{\it $\bullet$ Q2: From Fig.~\ref{young}b, there seems to be an unbalanced force component in the vertical direction $\gamma_{\rm LV}\sin\theta$. What force is missing to achieve equilibrium?}\\
{\it $\bullet$ Q3: Why does one draw a single force acting on the contact line in the case of the plate (Fig.~\ref{Young}a), while for Young's law we need to balance all three forces (Fig.~\ref{Young}b)?}

Actually, when measuring a surface tension using the plate technique, one often uses a platinum plate to be sure that the liquid completely wets the solid. In that case, however, $\gamma_{\rm SV}-\gamma_{\rm SL}>\gamma_{\rm LV}$ and Young's law does not apply! Then, the thermodynamic and mechanical approaches give conflicting answers:\\
{\it $\bullet$ Q4: In the situation of complete wetting, is the force on the plate given by $\gamma_{\rm LV}$ or by $\gamma_{\rm SV}-\gamma_{\rm SL}$?}

\subsection{Answers}

Before addressing these points in detail, we start with a short overview of the answers to the questions raised above. We emphasize that the thermodynamic result (i.e. from the virtual work principle) always gives the correct total force. If one wants to know the \emph{local} force distribution, which cannot be extracted from thermodynamics, it is imperative that the system on which the forces act is properly defined. Confusion regarding the forces is often caused by an improper or incomplete definition of such a system. \\
{\it $\bullet$ A1: The schematic of Fig.~\ref{MissingBounds} represents only the attractive intermolecular forces. The real force balance requires both repulsive and attractive interactions between liquid molecules.}

To answer the questions related to the contact line it is crucial to specify the system of molecules on which the forces are acting: \\
{\it $\bullet$ A2: In Young's law, the system \emph{on which the forces act} is a corner of liquid bounded by the contact line. $\gamma_{\rm LV}$ is indeed the force exerted on this system inside the liquid-vapor interface, but the forces exerted by the solid on the corner are incomplete in Fig.~\ref{Young}b. An extra vertical force on the liquid, caused by the attraction of the solid, exactly balances the upward force $\gamma\sin\theta$.\\}
{\it $\bullet$ A3: To obtain the force on the plate, the system to consider is the solid plate. In this case, the force exerted by the liquid on the solid is in fact equal to $\gamma_{\rm LV}\cos\theta$ per unit length. \\}
{\it $\bullet$ A4: The correct vertical force on the plate is $W~\gamma_{\rm LV} \cos \theta$. In the case of complete wetting ($\theta=0$), the virtual work principle can be applied, but only when taking into account the prewetting film.}

\section{Microscopic interpretation of capillarity}

\subsection{The liquid state}

\begin{figure}[t!]
\begin{center}
\includegraphics{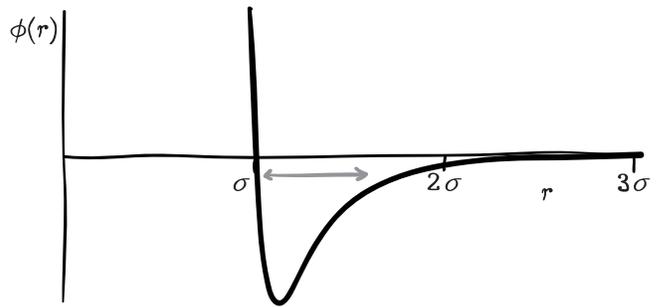}
\caption{\label{LJ}Lennard-Jones intermolecular potential $\phi$. The interaction is strongly repulsive for intermolecular distances $r < \sigma$, where $\sigma$ reflects the `hard-core' of the molecules. At large distances the molecules are attracted to one another. The gray arrow points to the presence of thermal fluctuations, which, in a liquid, lead to substantial variations of the intermolecular distance.}
\end{center}
\end{figure}

To address the origin of capillarity, we first have to understand how a liquid phase and a vapor phase can coexist. This goes back to the work of van der Waals who derived an equation of state for fluids that can account for a liquid-gas phase transition\cite{Rowlinson/Widom,Hansen}:
\begin{equation}
P=\frac{kT}{v-b}-\frac{a}{v^2}\;.
\label{VDW}
\end{equation}
Here $P$ is the pressure, $v$ the volume per molecule, and $T$ the temperature. This equation of state corrects the ideal gas law to incorporate the effect of intermolecular forces. The constant $b$ introduces repulsion between molecules as an excluded volume effect: the pressure diverges when the total volume per molecule reaches a minimal size $b$. In this limit the molecules are densely packed and constitute a liquid phase. In this phase, the volume per molecule no longer depends on pressure, which means that the liquid phase is incompressible. Ultimately, this effect comes from the repulsion of the electron clouds of the molecules, due to Pauli exclusion principle in quantum mechanics. The constant $a$ represents the long-range attraction between molecules which finds its origin in the dipole-dipole interaction (van der Waals attraction). 

Van der Waals's equation of state (\ref{VDW}) explains how a low density gas phase can coexist with a high density liquid phase. This coexistence requires the pressures to be identical on both sides of the interface, despite the striking difference in density. In a gas, where $v=v_g$ is large, most of the energy is of kinetic origin ($a/v_{g}^2\ll P$): the pressure is $P\simeq kT/v_{g}$. In the liquid phase the volume per unit molecule is almost in the incompressible limit: $v=v_{l}\approx b$. This strong repulsive effect ($kT/(v_{l}-b)\gg P$), however, is counterbalanced by the presence of attractions ($a/v_{l}^2\gg P$) so that, for the same temperature, the pressure in the liquid phase can be in equilibrium with the pressure in the gas phase. This gives rise to a stable liquid-vapor interface.

How can a liquid at the same time be repulsive and attractive? A single pair of atoms can of course only attract or repel each other, depending on the distance separating their two nuclei. This interaction is shown schematically in Fig.~\ref{LJ}. The steep potential reflects the short-range repulsion, while the negative tail of the potential represents the long-range attraction. This means that the balance of attractive and repulsive interactions in (\ref{VDW}) only has a statistical meaning: some particles are in an attractive state while others are in a repulsive state. This property of `simultaneous' repulsion and attraction is what makes a liquid very different from a solid: in an elastic solid, molecules in the same region are either all compressed (if the solid is submitted to an external compression), or all attracted to each other (if the solid is submitted to an external tension). This point is briefly discussed in the Appendix.

The difference between a solid and a liquid can be traced back to the importance of thermal fluctuations, i.e. the kinetic energy of the molecules. In the solid phase, these fluctuations are relatively small with respect to the potential energy, i.e. $k_B T \ll \epsilon$, where $\epsilon$ is the energy scale for the intermolecular forces. As a consequence the system only explores a small region of the potential. Hence the solid is either in the compressed or in the tensile state. On the other hand, the liquid phase is characterized by large fluctuations, for which $k_B T \sim \epsilon$. A broad range of the potential is therefore sampled by molecules in the same region of space (Fig.~\ref{LJ}). Finally, the case $k_B T \gg \epsilon$ corresponds to a gas phase of weakly interacting particles that is dominated by kinetic energy.

\subsection{The liquid-vapor interface: Question 1}
\subsubsection{The force of surface tension}

\begin{figure}[t!]
\begin{center}
\includegraphics{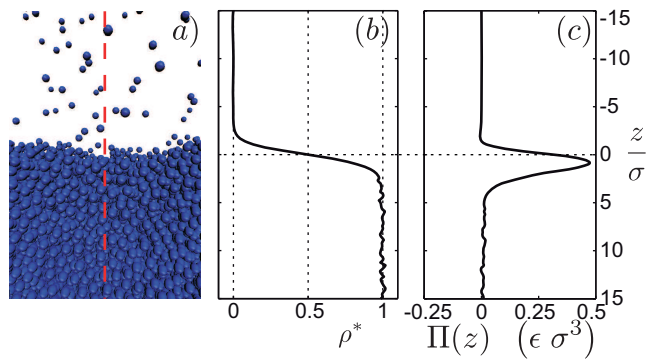}
\caption{\label{liquidvapor} The liquid-vapor interface. The vertical axis is in units of the molecular scale $\sigma$. (a) Snapshot of Lennard-Jones simulation of a liquid-vapor interface. (b) Time-averaged normalized density profile $\rho^\ast(z)$ across the interface. (c) Tangential force per unit area exerted by the left part on the right part of the system. Technically speaking, the plot shows the difference $\Pi=p_{NN}-p_{TT}$ between the normal and tangential components of the stress tensor.}
\end{center}
\end{figure}

Let us now consider the liquid-vapor interface in more detail. Figure~\ref{liquidvapor}a shows a snapshot of the interface obtained in Molecular Dynamics simulations of molecules that interact through a Lennard-Jones potential~\cite{Indekeu,Nijmeijer,Weijs}. The corresponding time-averaged density profile is plotted in Fig.~\ref{liquidvapor}b. The transition from the high density liquid to the low density gas takes place in a very narrow region that is a few molecules wide. In order to determine the capillary forces we need to divide the system along the direction normal to the interface into two subsystems (see Fig. \ref{liquidvapor}a). One considers the force per unit area, called the stress, exerted by the left subsystem on the right subsystem as a function of the vertical position $z$. This stress can be decomposed into two contributions: the pressure $P$, which we recall to be the same in the vapor and the liquid bulk, plus an extra stress $\Pi(z)$ acting along the direction parallel to the interface, plotted in Fig.~\ref{liquidvapor}c. The profile of this stress anisotropy shows that there is a force localized at the interface, acting in the direction parallel to the interface. This force spreads over a few molecular scales, which is also the typical thickness of the density jump across the interface. The integrated contribution of this force is indeed equal to $\gamma_{\rm LV}$ per unit length, the surface tension. This shows that surface tension really is a mechanical force.

\begin{figure}[t!]
\begin{center}
\includegraphics{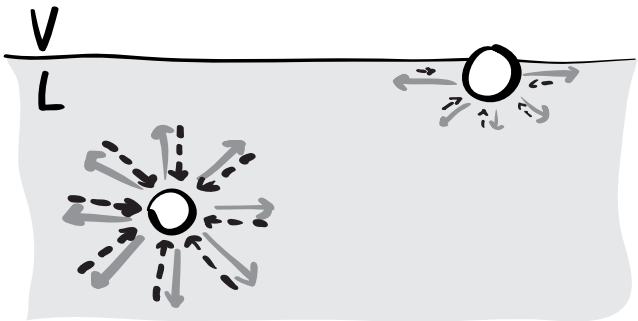}
\caption{\label{MissingBoundsCorrect}Sketch showing repulsive (dashed black arrows) and attractive (gray arrows) forces in the bulk and at the surface.}
\end{center}
\end{figure}

Having seen that in our simulations there is a parallel force localized at the interface, let us turn to Question 1. Why is the tension force parallel and not normal to the interface? First, we note that Fig.~\ref{MissingBounds} only depicts the attraction between molecules. A more complete picture also incorporates the repulsive forces in the internal pressure, as sketched as dashed arrows in Fig.~\ref{MissingBoundsCorrect}. Away from the surface there is perfect force balance due to the symmetry around the molecule. Near the interface, however, the up-down symmetry is broken. To restore the force balance in the vertical direction, the upward repulsive arrow (dashed) has to balance with the downward attractive arrow (solid). In the direction parallel to the interface the symmetry is still intact, thus automatically ensuring a force balance parallel to the interface. This means that along the direction parallel to the interface, there is no reason why the attractive forces should have the same magnitude as the repulsive forces. As described above, we find that in practice, the attractive forces are stronger, which indeed give rise to a positive surface tension force.

\subsubsection{Separate roles of attraction and repulsion}

However, we still need to explain why these intermolecular forces give rise to such a strong tension along the surface. This question was previously addressed by Berry\cite{Berry}, who noted the separate roles of attraction and repulsion. The key observation is that, to a good approximation, the repulsive contribution to the pressure is isotropic while attraction is strongly anisotropic. This is because the repulsion is very short ranged due to the hard core of the molecules, and can therefore be thought of as a ``contact force". As such, repulsion is not very sensitive to the changes in molecular structure near the interface and is equally strong in all directions\cite{Weijs}. By contrast, the long-range nature of the attractive forces make them very susceptible for the molecular structure. This is the origin of the observed pressure anisotropy near the interface that generates the surface tension force. 

\begin{figure}[t!]
\begin{center}
\includegraphics{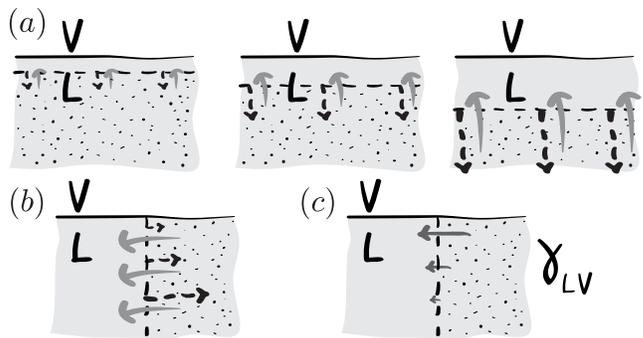}
\caption{\label{InterfaceLV} Forces exerted on a subsystem of liquid (dotted region) by the rest of the liquid (gray region without dots) (a) The subsystem considered is the lower part of the liquid and is separated from the interfacial zone by a line parallel to the liquid-vapor interface. The subsystem (dotted region) is submitted to an attractive force (gray arrows) and a repulsive force (dashed black arrows) exerted by the rest of the liquid (gray region without dots). They must balance each other. (b) The liquid is now divided along a line perpendicular to the interface. The subsystem considered (dotted region on the right) is submitted to an attractive force (gray arrows) and to a repulsive force (dashed arrows) exerted by the rest of the liquid (gray region without dots).As the repulsive force is isotropic, it has the same magnitude as in (a) and therefore decays close to the surface. On the contrary, the attractive force is nearly constant and remains almost unchanged close to the surface. (c) This leads to a net attractive force from one side on the other.}
\end{center}
\end{figure}

To see how this works out in detail it is useful to divide the liquid into two subsystems using control surfaces parallel to the liquid-vapor interface, as shown in Fig.~\ref{InterfaceLV}a. The force exerted on the dotted subsystem by the rest of the liquid results from the superposition of attractive (vertical gray arrows) and repulsive (dashed black arrows) interactions (Fig.~\ref{InterfaceLV}a). As the subsystem is at equilibrium, these attractive and repulsive components must balance each other. The magnitude of the attractive force increases with the size of attracting region -- this is because the density increases as one moves from the vapor towards the liquid phase. The magnitude of the attraction saturates to the bulk value when the control surface is a few molecular sizes from the interface. We then divide the liquid into two subsystems using a control surface perpendicular to the liquid vapor interface (Fig.~\ref{InterfaceLV}b). One can now use that the repulsive short-range forces are isotropic. This means that the magnitude of repulsion (dashed black arrows) exerted by the left side on the subsystem (dotted region) increases with depth in a way analogous to that in Fig.~\ref{InterfaceLV}a. By contrast, the strength of attraction has a much weaker dependence on depth -- for the sake of simplicity we draw it at a constant magnitude that equals the attraction in the bulk. As a result, there indeed is a net attraction of the subsystem by the rest of the liquid (dark gray arrow in Fig.~\ref{InterfaceLV}c).

\subsection{The liquid-solid interface}
\label{sect:LS}
\subsubsection{Forces near the liquid-solid interface}
\begin{figure}[t!]
\begin{center}
\includegraphics{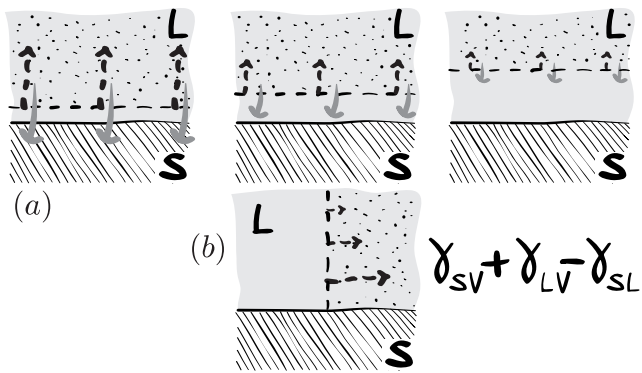}
\caption{\label{InterfaceLS}Forces exerted by the solid (dashed line) on a subsystem of liquid (dotted region). On these schematics, the attractive liquid-liquid interactions already treated in Fig.~\ref{InterfaceLV}, are not considered. (a) The liquid subsystem is semi-infinite. It is delimited by a line parallel to the liquid-solid interface, at different distances above it. The subsystem (dotted region) is submitted to an attractive force (gray arrows) exerted by the solid and to a repulsive force (dashed arrows) exerted by the rest of the liquid (gray region without dots). As the subsystem is in equilibrium, they must balance each other. (b) The liquid is now divided along a line perpendicular to the interface. Only the horizontal force components are shown. The solid exerts no horizontal attraction. As the repulsive interactions are isotropic, this results into an horizontal repulsive force exerted by one side on the other.}
\end{center}
\end{figure}

We now consider the liquid-solid interface (see Fig.~\ref{InterfaceLS}). Here, two effects superimpose. First, due to the presence of the solid, there is a lack of liquid in the lower half-space (hatched region in Fig.~\ref{InterfaceLS}). This missing liquid induces an anisotropy of the attractive liquid-liquid force in the same way as it does in the case of the liquid-vapor interface. Therefore, as in Fig.~\ref{InterfaceLV}c, the left hand side of the liquid exerts a net attractive force $\gamma_{\rm LV}$ per unit length on the right hand side subsystem. The second effect is due to the liquid-solid interaction. In the same way as with the liquid-vapor interface, we divide the liquid into two subsystems using a control surface parallel to the interface, as shown in Fig.~\ref{InterfaceLS}a. The attraction by the solid (gray arrow) decreases with the distance and is perfectly balanced by a short range liquid-liquid repulsive force (dashed arrows). Now, we divide the liquid into two subsystems using a control surface \emph{perpendicular} to the liquid solid interface (Fig.~\ref{InterfaceLS}b). Assuming once more that the liquid-liquid repulsion is isotropic, the left part of the liquid exerts a net repulsive force on the right subsystem (dotted region). This force is induced by the influence of the solid and, perhaps surprisingly, it is not equal to $\gamma_{\rm SL}$. Instead, it has been shown\cite{Nijmeijer} that this force is equal to $\gamma_{\rm SV}+\gamma_{\rm LV}-\gamma_{\rm SL}$, and this will be motivated in more detail below. 

To combine these two effects, we subtract the unbalanced attractive force due to the absent liquid ($\gamma_{\rm LV}$) from the repulsive force due to the solid ($\gamma_{\rm SV}+\gamma_{\rm LV}-\gamma_{\rm SL}$) to find the net repulsive force between the subsystems: $\gamma_{\rm SV}-\gamma_{\rm SL}$.

\subsubsection{Solid-liquid interaction and the surface tensions}

\begin{figure}[t!]
\begin{center}
\includegraphics{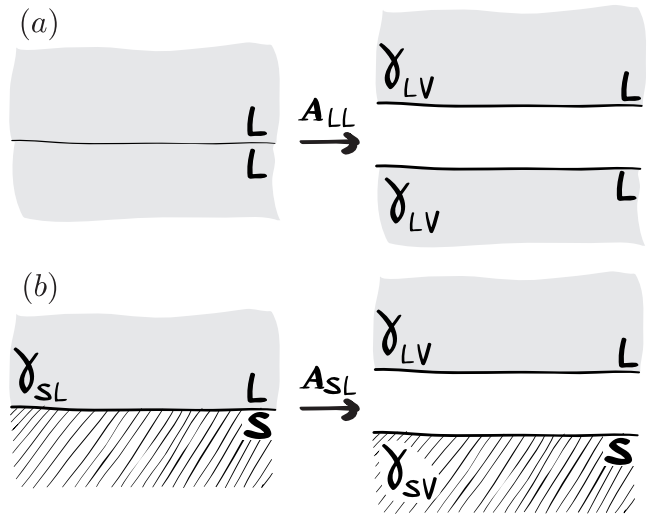}
\caption{\label{SL} Relation between adhesion work and surface tensions. (a) To split a liquid volume into two semi-infinite volume, one has to create two liquid-vacuum interfaces. It costs an energy $A_{\rm LL}=2\gamma_{\rm LV}$. (b) To create a solid-liquid interface one first needs to create a liquid-vacuum and solid-vacuum interface, which costs an energy $\gamma_{\rm SV} + \gamma_{\rm LV}$. Joining the liquid-vacuum and solid-vacuum interfaces yields an energy reduction $A_{\rm SL}=\gamma_{\rm SV} + \gamma_{\rm LV} - \gamma_{\rm SL}$ due to the solid-liquid attraction.}
\end{center}
\end{figure}

Let us motivate why the strength of the solid-liquid interaction does not couple directly to the solid-liquid surface tension $\gamma_{\rm SL}$, but to the combination $\gamma_{\rm SV}+\gamma_{\rm LV}-\gamma_{\rm SL}$\cite{deGennes,I91}. Note that this feature will also be crucial for understanding the wetting phenomena discussed in the following section.

The solid-liquid surface tension represents the free energy needed to create a solid-liquid interface. To make such an interface, one first has to ``break" a bulk solid and a bulk liquid into two separate parts, and then join these solid and liquid parts together. This breaking of liquid is depicted schematically in Fig.~\ref{SL}a (it works similarly for the solid). The corresponding energy is the ``work of adhesion" $A_{\rm LL}$ due to liquid-liquid attractions ($A_{\rm SS}$ for the solid). This gives rise to a surface tension $2\gamma_{\rm LV}=A_{\rm LL}$ (2$\gamma_{\rm SV}=A_{\rm SS}$), since the liquid (solid) is connected to a vacuum at this intermediate stage. When joining the solid-vacuum and liquid-vacuum interfaces, the attractive solid-liquid interaction will reduce the surface energy by the solid-liquid work of adhesion $A_{\rm SL}$ (Fig.~\ref{SL}b). Hence, the resulting solid-liquid surface tension becomes

\begin{equation}
\gamma_{\rm SL} = \gamma_{\rm SV} + \gamma_{\rm LV} - A_{\rm SL}.
\end{equation}
From this one finds that, indeed, the strength of the solid-liquid adhesion reads

\begin{equation}
A_{\rm SL} = \gamma_{\rm SV} + \gamma_{\rm LV} - \gamma_{\rm SL} 
= \gamma_{\rm LV}(1+\cos \theta).
\end{equation}
In the last step we used Young's law for the equilibrium contact angle. As a consequence, the capillary forces induced by solid-liquid attractions will have a magnitude $A_{\rm SL}= \gamma_{\rm LV}(1+\cos \theta)$ and not $\gamma_{\rm SL}$.

\section{Microscopic interpretation of wetting}

The question of the force balance is even more intricate in the vicinity of the contact line, where the liquid-vapor interface meets a solid. It is crucial to note that \emph{the contact line itself does not represent any material}: it is a mathematical line that marks the separation between wetted and dry parts of the solid. The question ``What is the force on the contact line?" is thus ill-posed, since there are no molecules on which such a force would act. As a matter of fact, only a material system (a collection of matter) can be submitted to a force. Therefore, care should be taken to properly define the systems that play a role near the contact line: the liquid near the contact line and the solid underneath it. In the following sections we will show how a careful consideration of all the forces on the appropriate material systems will lead to proper force balances, consistent with the thermodynamic predictions.

All results and sketches provided in this section, some of which may appear counterintuitive, are backed up by a Density Functional Theory model for microscopic interactions~\cite{BauerDietrich,GettaDietrich} and molecular dynamics simulations.\cite{Weijs}

\subsection{Force on a liquid corner: Question 2}
\begin{figure}[t!]
\begin{center}
\includegraphics{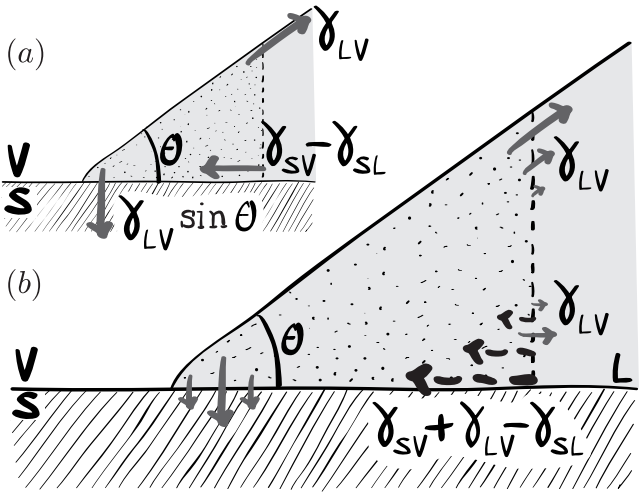}
\caption{\label{YoungCorrect} Solid and liquid forces acting on a liquid subsystem (dotted regions) near the contact line. (a) Sketch of a wedge of liquid near the contact line with the three forces exerted on the system. (b) Each of the three corners of this system must be treated differently. The upper right corner is at the liquid vapor interface. Following Fig.~\ref{InterfaceLV}, the rest of the liquid exert a net attractive force parallel to the interface equal to $\gamma_{\rm LV}$ per unit length. The lower right corner is at the liquid solid interface. Following Figs.~\ref{InterfaceLV} and \ref{InterfaceLS}, the rest of the liquid exerts a repulsive force $\gamma_{\rm SV}-\gamma_{\rm SL}$. The liquid near the solid-liquid interface is attracted by the solid, this force is balanced everywhere by repulsion at the solid-liquid interface, except in the vicinity of the contact line.}
\end{center}
\end{figure}
Let us consider the forces on the wedge-shaped liquid corner in the vicinity of the contact line, as shown in Fig.~\ref{YoungCorrect}. Choosing this subsystem, we will now explain Young's force construction from Fig.~\ref{Young}b and answer Question 2: What happens to the force balance normal to the solid-liquid interface?

\begin{figure}[t!]
\begin{center}
\includegraphics{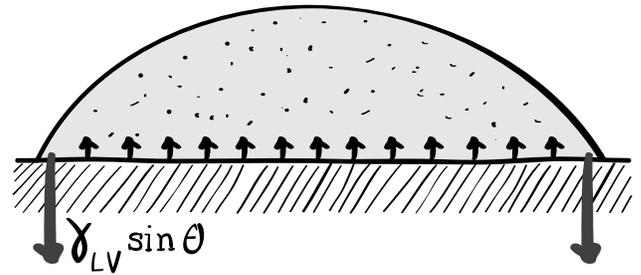}
\caption{\label{Laplace}Forces acting on a system, in this case a liquid drop (dotted area). The system is in equilibrium so the sum of all (external) forces must be zero. Due to Laplace pressure, there is a repulsive force exerted by the solid on the liquid across the liquid-solid interface (upward black arrows). In the vicinity of the contact line, repulsion and attraction of the liquid by the solid do not balance each other. Therefore, the solid attracts the liquid with a vertical force equal to $\gamma_{\rm LV}\,\sin\theta$ per unit length (downward dark gray arrows).}
\end{center}
\end{figure}

There are two types of forces acting on the liquid molecules inside the subsystem: interactions with the solid and interactions with other liquid molecules outside the subsystem. We first consider the solid-on-liquid forces. One can immediately see that, since the solid spans an infinite half space, every liquid molecule experiences a resultant force that is oriented purely normal to the solid-liquid interface: the left-right symmetry of the solid ensures that there is no force component parallel to the interface. Far from the contact line at the solid-liquid interface, this attractive force is balanced by a repulsive force, as discussed in Fig.~\ref{InterfaceLS}. However, since the repulsive force is continuous and zero outside the droplet, the repulsive force must decay close to the contact line. This means that there is an unbalanced attractive force that is strongly localized in the vicinity of the contact line. It has been shown\cite{SD2011} that this force per unit length is equal to $\gamma_{\rm LV}\,\sin\theta$. The existence of this force has recently been challenged\cite{Finn06,Finn08}. To show that this force indeed must exist to achieve equilibrium, we consider the droplet shown in Fig.~\ref{Laplace}. Choosing the drop as the system, and recognizing that the force in the interior of the droplet at the liquid-solid interface (small arrows) is due to the Laplace pressure $2\gamma_{\rm LV}\kappa$ (with $\kappa$ the curvature $1/R$), the attractive force at the contact line must be exactly $\gamma_{\rm LV}\sin\theta$ to achieve a force balance\cite{Rusanov75,Carre96,White03,Pericet-Camara2009,Pericet-Camara2009-2}. This provides the answer to Question 2: the downward solid-on-liquid force is not drawn in Fig.~\ref{Young}b. This missing force has often been interpreted as a reaction from the solid\cite{Quere}, whose existence is demonstrated experimentally by the elastic deformation of soft solids below the contact line\cite{Pericet-Camara2009,Pericet-Camara2009-2,Wang2009,Yu2009}. Here, we clarify the molecular origin of this normal force\cite{SD2011}.

To finalize the force construction near the contact line we return to the wedge shown in Fig.~\ref{YoungCorrect}b. Since the solid can only exert a normal force on the liquid, all parallel force components drawn in Young's construction are purely due to the liquid molecules outside the corner. The force drawn along the liquid-vapor interface can be understood directly from the tension $\gamma_{\rm LV}$ inside the liquid-vapor interface (cf.~the discussion of Fig.~\ref{InterfaceLV}). A similar force arises at the solid-liquid interface (Fig.~\ref{InterfaceLS}), which, however, is repulsive and has a magnitude $\gamma_{\rm SV}-\gamma_{\rm SL}$, as explained in the previous section. Including these forces gives a perfect force balance on the liquid corner, as seen in Fig.~\ref{YoungCorrect}a. In fact, one can easily verify that even the resultant torque (or force moment) is zero for this force construction. As such, it provides a more physical alternative to the classical picture of Young's law. 

\subsection{Liquid-on-Solid force: Question 3}

The measurement of surface tension in Fig.~\ref{Young}a directly relies on the force exerted by the liquid on the solid plate. Again, we emphasize the importance of a proper definition of ``the system'' on which the forces act, and in this case this is the solid on which the liquid rests. The situation is thus very different from the forces on the liquid corner, which are in equilibrium and experiences a zero resultant force. This \emph{difference between systems} provides the key to Question 3. In Fig.~\ref{Young}a, the total force exerted by the liquid on the solid is represented by the resultant $\vec\gamma_{\rm LV}$, while Fig.~\ref{Young}b represents the balance of the forces acting on the liquid wedge.

\subsubsection{Forces near the contact line}
\begin{figure}[t!]
\begin{center}
\includegraphics{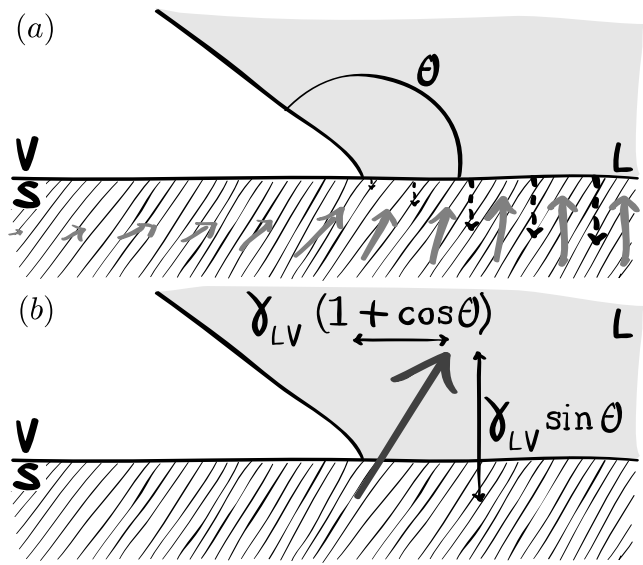}
\caption{\label{ForceSolid}
Forces acting on the solid subsystem (hatched areas) by the liquid (gray areas) near the contact line.
(a) Distribution of forces acting on the solid near the contact line. Due to the attraction of the liquid, the solid is attracted towards the liquid (solid gray arrows). The absence of liquid on the left part of the contact line ensures the tangential force is always towards the liquid, even for $\theta>90^\circ$.
The repulsion (dotted arrows) arises from the contact force at the solid-liquid interface.
Far from the contact line, repulsion and attraction balance each other.
(b) Resultant force acting on the solid near the contact line. The net normal force is $\gamma_{\rm LV}\sin \theta$, and the parallel force $\gamma_{\rm LV}+\gamma_{\rm SV}-\gamma_{\rm SL}=\gamma_{\rm LV}(1+\cos\theta)$.}
\end{center}
\end{figure}

Once again, we turn to the microscopic description of the forces in the vicinity of the contact line. It turns out that the normal component of the force exerted on the solid is equal to $\gamma_{\rm LV}\sin \theta$, consistent with the macroscopic picture of a tension along the liquid-vapor interface. However, the parallel component of the liquid-on-solid force does not have the expected magnitude $\gamma_{\rm LV}\cos \theta$, but $\gamma_{\rm LV}+\gamma_{\rm SV}-\gamma_{\rm SL}=\gamma_{\rm LV}\;(1+\cos\theta)$. This can be understood as follows. Fig.~\ref{ForceSolid}a illustrates that the tangential force component originates from the long-range attraction between solid and liquid molecules. In the previous section we already demonstrated the strength of this solid-liquid adhesion to be $A_{\rm SL}= \gamma_{\rm LV}\;(1+\cos\theta)$. Hence, there is no reason why the total force on the solid should be $\gamma_{\rm LV}\cos \theta$. Indeed, the DFT calculation confirms a tangential liquid-on-solid force of magnitude $A_{\rm SL}= \gamma_{\rm LV}\;(1+\cos\theta)$\cite{SD2011}.

The physics of this surprising result is nicely illustrated by Fig.~\ref{ForceSolid}. The macroscopic intuition that the resultant surface tension force is pulling along the liquid-vapor interface would predict a force to the left whenever the contact angle $\theta > 90^\circ$. However, it is clear from the sketch of the attractive forces that the sum of all parallel components must be oriented towards the liquid (right side in the figure). This stems from the asymmetry between the amount of liquid attracting the solid molecules on both sides of the contact line: there are always more liquid molecules on the right side of the contact line in this figure. This is consistent with a parallel force $\gamma_{\rm LV}\;(1+\cos\theta)$, but not with a force $\gamma_{\rm LV} \cos \theta$ (which changes sign at $90^\circ$). Note that when considering the force exerted by the solid \emph{on the liquid}, this asymmetry does not occur since the solid is left-right symmetric, therefore there is no tangential component. This once more illustrates that a detailed force interpretation crucially relies on the definition of the system.
\begin{figure}[t!]
\begin{center}
\includegraphics{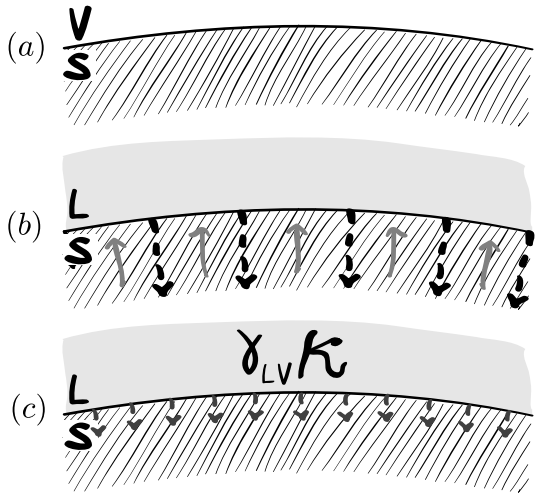}
\caption{\label{LaplaceSolid} Forces acting on a solid at the solid-liquid interface. (a) Without liquid, there is neither repulsion nor attraction. (b) When liquid is present there is repulsion and attraction. However, the repulsion is not completely balanced in this curved interface, as there is more liquid than in a plane geometry. (c) The resulting force per unit surface is $\gamma_{\rm LV}\kappa$ where $\kappa$ is the curvature of the liquid (positive in this case), equivalently to the Laplace pressure. It only shows a dependency on the liquid-liquid interactions as it is only the absence of liquid that is relevant to the curvature.}
\end{center}
\end{figure}

\subsubsection{Global force balance: curvature of solid-liquid interface}

To solve this apparent discrepancy with the thermodynamic result one has to consider all the forces exerted by the liquid on the solid, i.e. not just the forces near the contact line. The key point is that the submerged solid bodies cannot be flat everywhere and must have curved pieces of liquid-solid interface. If the interface separating the solid from the liquid is flat, the net normal force is locally zero as repulsion balances attraction (far away from the contact line). However, when this interface is curved, a repulsive force inside the liquid is enhanced due to the curvature of the solid-liquid interface, in a way similar to the Laplace pressure. As shown in Fig.~\ref{LaplaceSolid}, the presence of a curved half-space of liquid acts on the solid, and creates an unbalanced liquid-on-solid force $\gamma_{\rm LV}\kappa$ per unit area. Here $\kappa$ is the local curvature (inverse radius of curvature) of the solid. Interestingly, DFT calculations\cite{SD2011} show that the resultant pressure couples only to $\gamma_{\rm LV}$ and not to $\gamma_{\rm SL}$. As we show below, this is exactly what is needed to restore consistency between microscopic and thermodynamic forces. 

\begin{figure}[t!]
\begin{center}
\includegraphics{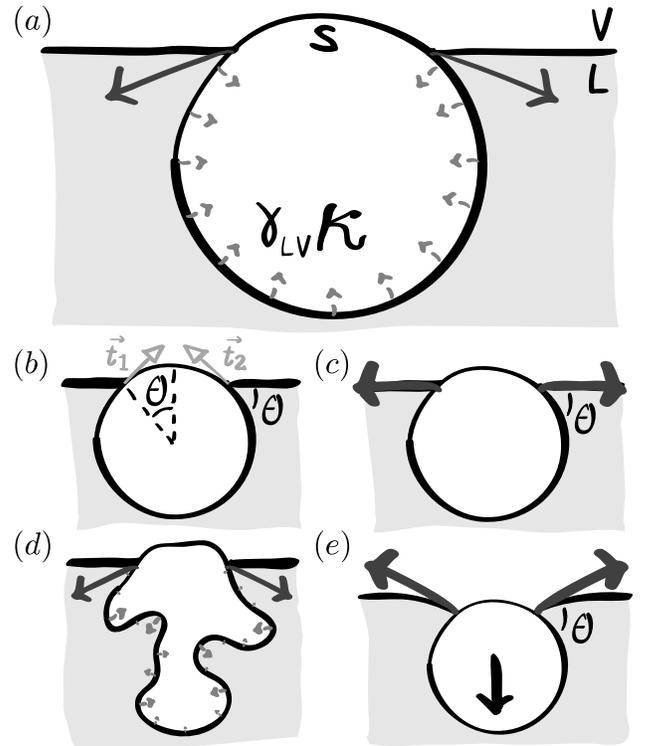}
\caption{\label{Pin}Capillary forces on immersed objects. (a) Schematic of a pin floating at the surface of a liquid under partial wetting conditions and in zero-gravity. The downward thin gray arrows are the forces exerted by the liquid on the pin located in the vicinity of the contact line. The small light gray arrows show the Laplace pressure $\gamma_{\rm LV}\;\kappa$ acting on the solid, due to the curvature $\kappa$ of the solid-liquid interface. (b) $\theta$ denotes the contact angle and $\vec t_{1}$ and $\vec t_{2}$ are two unit vectors tangential to the pin, pointing upwards, at the two contact lines. (c) The thick gray arrows show the resultants of the capillary forces detailed in (a), which apply on each half of the pin. They reduce to forces tangential to the liquid-vapor interface at the contact lines. Note that this schematic does not show the distribution of capillary forces. (d) Distribution of the capillary forces in the case of an irregular shape. Since the integral over the curvature is equal to the sum of the two tangential vectors at the contact lines, the resultant is independent of the shape of the body. It is thus the same as in (c). (e) Pin floating at the surface of a liquid under gravity. The upward thick gray arrows are the resultants of capillary forces. They balance the effect of gravity (corrected by the Archimedes force), shown as the downward black arrow.}
\end{center}
\end{figure}

An excellent demonstration of this effect is the long debated case of a `floating-pin' under zero-gravity, as shown in Fig~\ref{Pin}. Whereas a floating pin in a system with gravity leaves a visible depression in the liquid-vapor interface near the contact line (cf.~Fig.~\ref{Pin}e), the zero-gravity condition ensures that the interface has a constant curvature, i.e. it is straight everywhere. This means that the vertical position depends on the equilibrium contact-angle alone, and not on the density ratio of the involved materials. As discussed in Fig.~\ref{ForceSolid}, the liquid-on-solid force near the contact line is not oriented along the liquid-vapor interface, but points towards the interior of the liquid. This creates a resultant downward force that has to be compensated to restore equilibrium. Additionally, the curvature of the solid-liquid interface creates a normal force distributed over the whole immersed surface of the solid of magnitude $\gamma_{\rm LV}\kappa$ per unit surface (cf.~Fig.~\ref{Pin}a). Integrating over the curvature of the submerged surface from one contact line to the other then gives the resultant of the Laplace pressure:
\begin{align}
\gamma_{\rm LV}\int_1^2\kappa\vec n dS = \gamma_{\rm LV}(\vec t_2 + \vec t_1)\;,
\end{align}
where $\vec t_{1}$ and $\vec t_{2}$ are the unit vectors tangential to the pin, pointing upwards. Therefore, the resultant is orientated upwards and is equal to $2 \gamma_{\rm LV}\sin \theta$ per unit depth (see Fig.~\ref{Pin}b). It balances exactly the downward forces induced close to the contact lines (Fig.~\ref{ForceSolid}b), hence the pin is in equilibrium. Importantly, this result is independent of the shape of the body (see Fig.~\ref{Pin}d).

The same principle applies to the partially wetted plate of Fig.~\ref{Young}a: the force exerted by the fluid on the plate results from two contributions, as shown schematically in Fig.~\ref{PlatePartialTotal}c. First, there is the vertical force component (per unit length) due to the vicinity of the contact line: $\gamma(1+\cos\theta)$ (cf. Fig.~\ref{ForceSolid}). Second, there are submerged surfaces of the plate where a localized curvature exists at the corners. This curvature induces a Laplace force on the pin (see Fig.~\ref{Pin}d) which results into a net upward force $\gamma_{\rm LV}$ per unit length of contact line which means the total force (per unit length of contact line) on the plate is $\gamma_{\rm LV} \cos \theta$, in agreement with the thermodynamic result.

\subsection{Complete wetting: Question 4}

\begin{figure}[t!]
\begin{center}
\includegraphics{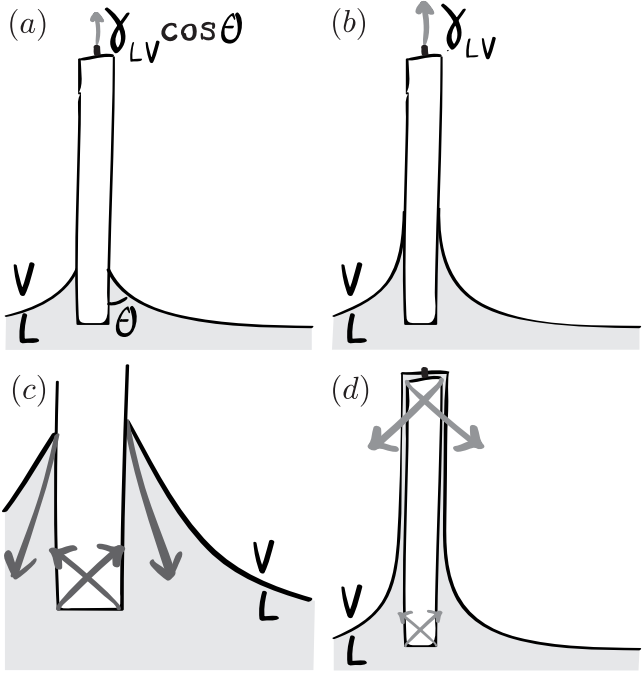}
\caption{\label{PlatePartialTotal} (a-b) Force per unit length of the contact line needed to keep a plate in equilibrium in a bath in partial (a) and complete wetting (b). (c) Partial wetting. The vertical force at the contact line, equal to $\gamma_{\rm LV}(1+\cos\theta)$, is balanced by the Laplace pressure induced by the curvature of the plate. Note that any plate shape would lead to the same resultant force as the integral of the curvature over the surface reduces to the local tangents at the contact line (see Fig.~\ref{Pin}d). (d) Complete wetting case. Due to the mesoscopic pre-wetting film, of which the thickness is exaggerated in the picture, there is no contact line, thus there is no force located near the apparent macroscopic contact line. The forces are related only to the Laplace pressure: the curvature of the solid gives a zero resultant force, as it is completely immersed in the liquid, and the curvature of the liquid is acting on the solid only in the pre-wetting zone --- as it is compensated by gravity in the meniscus. The resultant force per unit length is then equal to $\gamma_{\rm LV}$.}
\end{center}
\end{figure}

In the case of complete wetting, Young's law for the contact angle is no longer applicable. Instead, the apparent contact angle $\theta$ vanishes because the three surface tensions do not balance each other:
\begin{equation}
\gamma_{\rm SV}-\gamma_{\rm SL}>\gamma_{\rm LV}\;.
\end{equation}
Physically, there is no real contact line in this configuration (Fig.~\ref{PlatePartialTotal}b), but there is a meniscus where the liquid-vapor interface approaches the solid. Beyond the meniscus, there exists a mesoscopic liquid film called a pre-wetting film, which covers the solid completely (Fig.~\ref{PlatePartialTotal}d). The existence of an \emph{apparent} contact line is only due to the effect of gravity: on a flat surface, the liquid would simply spread. The interface between the liquid and vapor phases consequently presents two regions. In the lower region, the meniscus can be described by the balance between the Laplace pressure and the gravity potential 
\begin{equation}
\gamma_{\rm LV} \kappa=\rho g z\;,
\end{equation}
where $z$ is the height above the bath (thus, no additional constant needed) and $\kappa$ the curvature of the interface. Introducing the capillary length $\ell_\gamma=\sqrt{\gamma_{\rm LV} /\rho g}$ this equation can be written as: $\ell_\gamma^2 \kappa=z$. In the upper region, there is the pre-wetting film whose thickness $h(z)$ is determined by the balance between the gravity potential and the disjoining pressure $\Pi(h)$. The disjoining pressure is by definition the potential energy per unit volume at the surface of a liquid layer of thickness $h$, therefore the balance reads\cite{deGennes,I91}:
\begin{equation}
\label{PrewetPressureBalance}
\Pi(h)=\rho g z\;.
\end{equation}
As the pre-wetting film is very flat, the contribution of Laplace pressure can be neglected ($\kappa=0$) in this regime. The disjoining pressure scales as:
\begin{equation*}
\Pi(h) \simeq \frac{\left(\gamma_{\rm SV}-\gamma_{\rm SL}-\gamma_{\rm LV}\right)\sigma^2}{h^3}\;,
\end{equation*}
for films where $h\gg \sigma$, where $\sigma$ is a length on the order of the molecular size. The disjoining pressure vanishes in the limit case $\gamma_{\rm SV}=\gamma_{\rm SL}+\gamma_{\rm LV}$, which can be interpreted as the situation for which the interaction is the same with the liquid and with the solid. Then, one indeed does not expect any influence of the thickness $h$ on the energy.

Equating gravity and disjoining pressure (eq. \ref{PrewetPressureBalance}), we obtain the thickness profile in the pre-wetting region:
\begin{equation*}
h(z)\simeq \left[\frac{\left(\gamma_{\rm SV}-\gamma_{\rm SL}-\gamma_{\rm LV}\right)\sigma^2}{\rho g z}\right]^{1/3}
\end{equation*}
In the vicinity of the apparent contact line, where the two zones must match, the thickness is thus of order $l_\gamma^{1/3} \sigma^{2/3}$. As $l_\gamma$ is millimetric and $\sigma$ nanometric, $h$ is mesoscopic ($h\simeq 100$~nm). From the microscopic point of view, the solid is then completely surrounded by a semi-infinite layer of liquid ($h\gg\sigma$). Therefore, the only forces acting on a solid in complete wetting are normal contact forces, such as Laplace pressures. There are no contact line forces such as those described in Fig.~\ref{ForceSolid}b.

The forces exerted by the liquid on the solid are then related, as previously, to the curvature of the liquid-solid interface but also, inside the pre-wetting film, to the curvature of the liquid-vapor interface (Fig~\ref{PlatePartialTotal}d). Integrated over the whole submerged solid, the curvature of the solid gives a zero resultant force, whereas the curvature of the liquid is only integrated where the pre-wetting film exists. As a result, the resultant force is vertical and has an amplitude $\gamma_{\rm LV}$ per unit length of the apparent contact line.

This result is indeed consistent with the thermodynamic perspective. Since the solid is covered by a liquid layer much thicker than the molecular size the surface tension above the apparent contact line is not $\gamma_{\rm SV}$ but $\gamma_{\rm SL}+\gamma_{\rm LV}$: the plate is always completely submerged. In essence, this means that the plate never leaves the liquid bath when the plate is pulled upwards. When moving, there is no change of the solid-vapor interfacial area (it remains zero) or of the solid-liquid interfacial area (which is simply the total area of the plate). The only change occurs at the liquid-vapor interfacial area, which is increased, and the required pulling strength is thus $\gamma_{\rm LV}$ per unit length of the apparent contact line.

\section{Conclusion}

\subsection{Summary}

In this article we have raised simple questions about capillarity that many students face. By studying the interfaces from a microscopic perspective, we were able to provide answers to these questions, and also reconcile thermodynamics and statistical physics. 

We have provided a mechanical perspective about why there exists an attractive force parallel to interfaces: surface tension. The absence of liquid creates an attractive anisotropic force within a few molecular lengths from the interface whereas the repulsion remains isotropic and scales with the local density of the fluid. The attractive anisotropy leads to a strong localized stress parallel to the interface called surface tension. This occurs at liquid-solid interfaces as well, where there is also a half-space of liquid missing.

The problems when constructing force pictures in capillarity often arise from the improper definition of a \emph{system} on which the forces act. Considering a corner of liquid near the contact line as a system, we have proposed here an alternative to Young's construction (Fig.~\ref{Young}b). The analysis allows to locate and understand the different forces, in particular the attractive force exerted by the solid. This new force constructions leads to a perfect mechanical equilibrium: vectorial force balance and torque balance.

When looking at the force that is exerted by the liquid on the solid near the contact line we find that, surprisingly, this force is \emph{not} $\gamma_{\rm LV}\cos \theta$, but $\gamma_{\rm LV}(1+\cos \theta)$. Moreover, a normal stress is exerted in all the regions of any curved solid-liquid interface: the liquid pulls the solid when the latter is convex. This force is equivalent to the usual Laplace pressure. One has to take \emph{both} these forces into account, to obtain the net force from the thermodynamic result. The advantage of this microscopic force description is that it provides a simple answer to an academic problem which has become controversial in the last decade: the floating pin paradox\cite{Finn06,Lunati07,Finn08}.

The drawings and several relations presented in this article are based on results obtained using the Density Functional Theory in the sharp-kink approximation\cite{SD2011}, which allows to make quantitative predictions of the force distributions in the liquid and in the solid. .

\subsection{Teaching perspective}

We realize that the detailed picture of microscopic forces is not necessarily the most accessible for teaching purposes. In particular when introducing the basic concepts of capillarity, it is much simpler to work from the thermodynamic perspective: energy minimization naturally yields the equilibrium conditions, while the resultant forces can be computed from the virtual work principle. Nevertheless, our analysis provides a number of insights that are useful when teaching capillarity:

\begin{itemize}

\item To determine ``capillary forces" it is crucial to explicitly specify the system (= collection of matter) to which the forces are applied. 

\item The surface tensions $\gamma_{\rm SL}$ and $\gamma_{\rm SV}$ do not pull on the solid. 

\item The global force exerted on the solid by the liquid can be computed by adding the contributions of (Laplace) pressure inside the liquid and a localized tension force $\gamma_{\rm LV}$ parallel to the liquid-vapor interface. While this gives the correct answer, it does not reflect the true microscopic distribution of liquid-on-solid forces.

\item By contrast, the resultant force on the liquid near the contact line \emph{does} involve the three surface tensions $\gamma_{\rm LV}, \gamma_{\rm SL}$ and $\gamma_{\rm SV}$. 

\item The classical construction of Fig.~\ref{Young}b to explain Young's law does not accurately represent the force balance. A complete picture is provided in Fig.~\ref{YoungCorrect}a. 
\end{itemize}
We hope in particular that the force construction of Fig.~\ref{YoungCorrect}a will find its way to the classroom to explain Young's law (\ref{young}). It is conceptually simple, it clarifies the ``system" to which forces are applied, and represents a perfect mechanical equilibrium. Namely, besides a balance of normal and tangential components, the forces also exert a zero torque. 

Finally, we note that the virtual work principle yields the correct resultant force on a solid, but that it cannot recover the true microscopic force distribution. The knowledge of such a force distribution is crucial when one wants to take into account how a solid is elastically deformed by the contact line\cite{Rusanov75,Carre96,White03}. Even though these deformations can be as small as a few nanometers, these can be measured due to the advent of modern experimental techniques\cite{Pericet-Camara2009,Pericet-Camara2009-2,Wang2009,Yu2009}. This experimental access renews the fundamental interest in the microscopic details of capillarity\cite{SD2011}.

{\bf Acknowledgements:~}We thank J. van Honschoten and K. Winkels for critically reading the manuscript. B.A. thanks the students of the University Paris-Diderot's Master of Physics for both their impertinent and pertinent questions. The example developed in the appendix was brought to us by Y. Forterre.

\appendix
\section{On the difference between liquid vapor and solid vapor surface tensions}

\begin{figure}[b!]
\begin{center}
\includegraphics{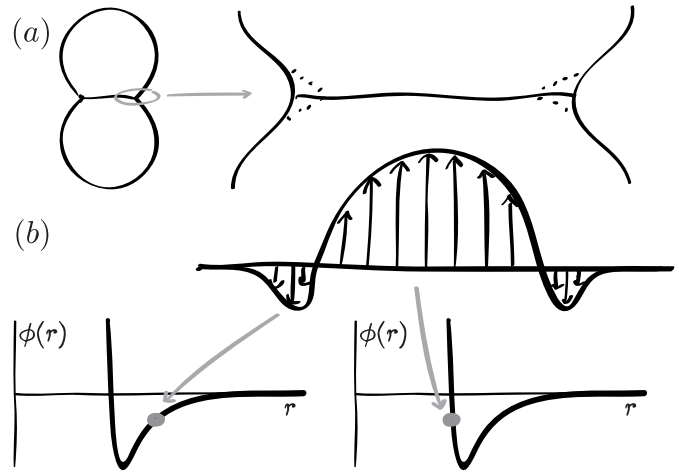}
\caption{\label{ElasticSpring} Unlike liquids, solids are locally either in a repulsive state or an attractive state. This is illustrated here by the contact between two adhesive solid elastic spheres~\cite{T77,G97,JG97,YCG07}. (a) Geometry of the contact in the presence (solid line) or absence (dashed line) of adhesion. (b) Radial distribution of normal force in the contact zone}
\end{center}
\end{figure}

From the mechanical point of view, the surface tensions of liquids and solids are fundamentally different. To illustrate this difference, let us analyse the example of Fig.~\ref{ElasticSpring}, which shows two small adhesive solid spheres such as grains in a powder of flour. One might imagine that the particles are attracted all over the contact, but in reality, there is also repulsion to give a zero resultant force. Although there is both attraction and repulsion in this example, the attractive and repulsive parts are spatially separated: attractive near the edge, repulsive in the central part of the contact. In a liquid, by contrast, the attraction and repulsion occur at the same location.

\end{document}